\documentclass[a4paper,twoside]{article}

\usepackage{epsfig}
\usepackage{subcaption}
\usepackage{calc}
\usepackage{amssymb}
\usepackage{amstext}
\usepackage{amsmath}
\usepackage{amsthm}
\usepackage{multirow}
\usepackage{pslatex}
\usepackage{apalike}
\usepackage{algorithm2e}
\usepackage[bottom]{footmisc}
\let\bibhang\relax
\usepackage[table]{xcolor}
\usepackage{natbib}

\usepackage{arydshln}

\usepackage{SCITEPRESS}     % Please add other packages that you may need BEFORE the SCITEPRESS.sty package.

\begin{document}
\newcommand{\textib}[1]{\textbf{\textit{#1}}}

\title{Efficient and Accurate Hyperspectral Image Demosaicing with Neural Network Architectures}

\author{\authorname{Eric L.~Wisotzky\sup{1,2}\orcidAuthor{0000-0001-5731-7058},
	Lara Wallburg\sup{1},
	Anna Hilsmann\sup{1} \orcidAuthor{0000-0002-2086-0951}, 
	Peter Eisert\sup{1,2} \orcidAuthor{0000-0001-8378-4805},
    Thomas Wittenberg\sup{3,4},
    Stephan G\"ob\sup{3,4} \orcidAuthor{0000-0002-1206-7478} }
\affiliation{\sup{1}Computer Vision \& Graphics, Fraunhofer HHI, Einsteinufer 37, 10587 Berlin, Germany}
\affiliation{\sup{2}Department of Informatics, Humboldt University, Berlin, Germany}
\affiliation{\sup{3}Fraunhofer IIS, Erlangen, Germany}
\affiliation{\sup{4}Chair of Visual Computing, Friedrich-Alexander-Universität Erlangen-Nürnberg, Erlangen, Germany}
\email{\{first\}.\{last\}@hhi.fraunhofer.de}
}


\abstract{Neural network architectures for image demosaicing have been become more and more complex. This results in long training periods of such deep networks and the size of the networks is huge. These two factors prevent practical implementation and usage of the networks in real-time platforms, which generally only have limited resources.
This study investigates the effectiveness of neural network architectures in hyperspectral image demosaicing. We introduce a range of network models and modifications, and compare them with classical interpolation methods and existing reference network approaches. The aim is to identify robust and efficient performing network architectures. 
Our evaluation is conducted on two datasets, "SimpleData" and "SimRealData," representing different degrees of realism in multispectral filter array (MSFA) data. The results indicate that our networks outperform or match reference models in both datasets demonstrating exceptional performance.
Notably, our approach focuses on achieving correct spectral reconstruction rather than just visual appeal, and this emphasis is supported by quantitative and qualitative assessments. Furthermore, our findings suggest that efficient demosaicing solutions, which require fewer parameters, are essential for practical applications. This research contributes valuable insights into hyperspectral imaging and its potential applications in various fields, including medical imaging.}

\onecolumn \maketitle \normalsize \setcounter{footnote}{0} \vfill
\section{\uppercase{Introduction}} \label{sec:introduction}
The use of multispectral images (MSIs) or hyperspectral images (HSIs), which encompass a wide range of different spectral channels across various wavelengths both within and beyond the visible spectrum, has gained increasing prominence in recent years. These types of images find broad applications in various fields such as healthcare \citep{calin2014hyperspectral,Lu2014,zhang2017tissue}, industrial applications \citep{shafri2012hyperspectral}, and agriculture \citep{jung2006hyperspectral,moghadam2017plant}. However, conventional acquisition methods and devices are associated with significant drawbacks, including high costs and lengthy acquisition times \citep{Wisotzky2018,WisotzkyComparision2021}.

In recent times, alternative approaches have been developed to address these challenges. One promising technique is based on the use of spectral masking at the pixel level, utilizing only a single sensor plane. This concept is known as Multi-Spectral Filter Arrays (MSFAs).
Unlike RGB images, which are composed of three color values (red, green and blue), MSFAs map the spectrum in more than three spectral bands, e.g., nine, 16 or 25 bands \citep{hershey2008multispectral}. A data cube is formed from the determined data. Its edges represent the image dimensions in x- and y-direction and the determined wavelengths in $\lambda$-direction. In contrast to MSIs, HSIs use several hundred spectral bands to capture the spectrum of a source.

When image data is acquired by a multispectral camera using the principle of MSFA, not all image information of a data cube can be acquired. An increase in spectral resolution is accompanied by a loss of spatial resolution. This missing information needs a successive interpolation using image processing or image analysis techniques. Common image processing techniques for spectral reconstruction include bilinear and nonlinear filters, referred to as debayering or demosaicing.

While MSFAs offer the advantage of real-time implementation and general applicability, accurate interpolation of missing spectral or spatial information is challenging. This, however, is crucial for precise spatial localization of entities such as cancer cells in healthcare, damage in plants in agriculture, or objects in industrial applications. To enhance existing methods, neural networks have recently been proposed as promising approaches.

\begin{figure}[htp]
  \centering
  \includegraphics[width=1\columnwidth]{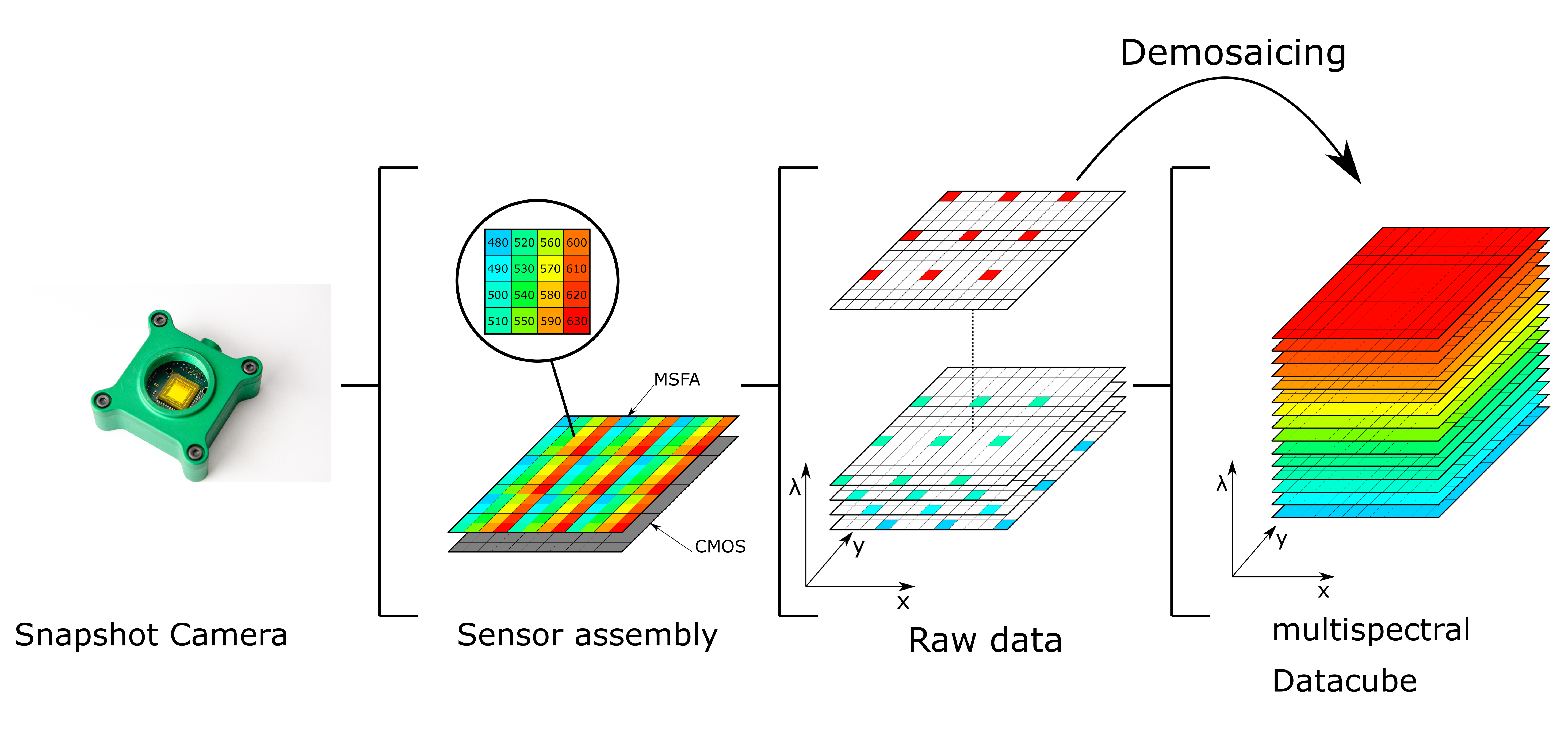}
  \caption{ Description of the processing pipeline and demosaicing of multispectral cameras.}
	\label{fig:processing_pipeline}
\end{figure}

The contribution of this work is as follows. We present a pipeline to achieve real-world MSFA data from different hyperspectral datasets. Further, we propose different new and extended demosaicing networks and compare these using classical public available datasets and a dataset processed with our introduced pipeline.

The analysis is performed from the perspective that the networks should provide good reconstruction quality and be able to compute it quickly, i.e., contain as few parameters as possible for potential near-sensor processing.

The remainder of this paper is as follows. The next chapter gives an overview on related publications relevant for this work. Chapter \ref{sec:network} describes the proposed network architectures, before chapter \ref{sec:data_training} explains data processing and presents training and evaluation parameters. Chapter \ref{sec:res} describes experiments and results, followed by a thorough discussion and conclusion.

%--------------------------------------------------------------%
\section{\uppercase{Related Work}} \label{sec:related_work}

The most commonly used algorithms for demosaicing in imaging are analytical ones such as nearest neighbor, bilinear interpolation or cubic interpolation. The multispectral debayering relies on these existing algorithms used in RGB imaging. The basic approach involves classic interpolation techniques. However, more complex methods take into account not only the information from the nearest pixel of the same channel but also the next of other channels or even further away pixels. There are many different variations of demosaicing algorithms available for RGB imaging~\citep{malvar2004high}. %For instance, Malvar et al.~\cite{malvar2004high} describes a highly qualitative method for reconstructing Bayer patterns using linear interpolation.
However, implementing these algorithms for multispectral filter structures is challenging due to the different number of channels.
An approach based on the weighted bilinear interpolation~\citep{brauers2006color} is the advancement of linear interpolation using MSFAs, called Intensity Difference (ID interpolation)~\citep{Mihoubi2015}.

Further, interpolation methods based on image fusion have been proposed~\citep{eismann2005hyperspectral,bendoumi2014hyperspectral,zhang2014spatial}. 
Fusion based methods usually require the availability of a guiding image with higher spatial resolution, which is difficult to obtain in many scenarios.

Demosaicing by interpolation based techniques, both traditional as well as fusion-based, is easy to achieve, however, these methods suffer from color artifacts and lead to lower spatial resolution. %Single pixel values are not exactly representative for the corresponding environmental information in the scene, i.e.~shapes and objects, 
Especially at edges, they do not take into account the spectral correlations between adjacent bands as well as due to crosstalk. This results in spectral distortions in the demosaiced image, especially for increasing mosaic filter size.

Alternatively, deep neural networks can be trained to account for scene information as well as correlations between individual spectral bands.

\subsection{Network-based Demosaicing}

Demosaicing using convolutional neural networks (CNN) was first proposed for images with $2\times2$ Bayer pattern~\citep{wang2014multilayer,gharbi2016deep}. In recent years, CNN based color image super resolution (SR) has gained popularity. Examples of such networks include SRCNN \citep{dong2014learning}, DCSCN \citep{yamanaka2017fast} and EDSR \citep{lim2017enhanced}. Due to their success, these networks have been extended to HSI super resolution \citep{li2018single}. The underlining aspect of all CNN based HSI demosaicing networks is the utilization of spatial and spectral context from the data during training.
%This flexibility and the trainability is a high advantage of network-based demosaicing.

Nevertheless there is only a small number of publications where CNNs are used for patterns $>2\times2$. In the NTIRE 2022 Spectral Demosaicing Challange several deep learning-based demosaicing algorithms were introduced \citep{arad2022ntire}. The leading methods all involved very large and complex network structures, such as an enhanced HAN \citep{niu2020single}, NLRAN and a Res2-Unet \citep{song2022hyperspectral} based method.
An interesting approach for the reconstruction of MSI uses leaner network structures of five residual blocks~\citep{shinoda}. Bilinear interpolated data are used as input for this refinement approach. Furthermore, there are already approaches to replace the bilinear interpolation by the ID interpolation \citep{gob2021multispectral}. \citet{dijkstra2019hyperspectral} proposed a similarity maximization network for HSI demosaicing, inspired by single image SR. This network learns to reconstruct a downsampled HSI by upscaling via deconvolutional layers. The network results are presented for a $4\times4$ mosaic pattern and the demosaiced HSI showed high spatial and spectral resolution. \citet{habtegebrial2019deep} used residual blocks to learn the mapping between low and high resolution HSI, inspired by the HCNN+ architecture \citep{shi2018hscnn+}.
These two networks use 2D convolutions in order to learn the spectral-spatial correlations. An important characteristic of HSI is the correlation between adjacent spectral bands which are not taken into account when using 2D convolutional based networks. These correlations can be incorporated by using 3D convolutional networks. \citet{mei2017hyperspectral} proposed one of the first 3D CNNs for hyperspectral image SR addressing both the spatial context between neighboring pixels as well as the spectral correlation between adjacent bands of the image. A more effective way to learn the spatial and spectral correlations is through a mixed 2D/3D convolutional network \citep{li2020mixed}. 

Further, deep learning (DL) have been employed in order to predict HSI data from MSI or even classic RGB image data~\citep{Arad2022Challenge,Arad_2020_CVPR_Workshops}. However, the main problem here is the lack of training data and the large dependence of the method on the application and training data. This means that the spectral behavior of individual scenes is learned by the DL model for interpolation and thus there is a dependency between the planned application and training data, which can lead to poor results and incorrect hyperspectral data if the training data is improperly selected, of poor quality, or too small in scope \citep{wang2021deep}.
Thus, the need of high quality ground truth data is essential. In such a dataset, each pixel should contain the entire spectral information, which is difficult to acquire in a natural environment.  

\subsection{Datasets}
\begin{table*}[ht]
\caption{The used hyperspectral data collections.}
\small
\centering
    \begin{tabular}{r|lllll}
    Dataset & \# Images & Size [px] & Spectrum [nm] & Bands & Range\\ \hline
    CAVE \citep{yasuma2010generalized} & 32 & 512x512 & 400-700 & 31 & 1:65536 \\
    HyTexiLa \citep{khan2018hytexila} & 112 & 1024x1024 & 405-996& 186 & 0:1\\
    TokyoTech31 \citep{monno2015practical} & 30 & 500x500 & 420-720 & 31 & 0:1\\
    TokyoTech59 \citep{monno2018single} & 16 & 512x512 & 420-1000 & 59 & 0:1\\
    SIDRI-v10 \citep{mirhashemi2019configuration} & 5 & 640x480 & 400-1000 & 121 & 0:1 \\
    SIDRI-vis \citep{mirhashemi2019configuration} & 1 & 640x480 & 400-720 & 31 & 0:1 \\
    ODSI-DB Nuance \citep{hyttinen2020oral} & 139 & 1392x1040 & 450-950 & 51 & 1:65536 \\
    ODSI-DB Specim \citep{hyttinen2020oral} & 171 & Various & 400-1000 & 204 & 1:65536 \\
    Google & 670 & 480x1312 & 400-1000 & 396 &  1:4096 \\
    \end{tabular}
    \label{tab:Databases}
\end{table*}
In order to be able to train and evaluate networks, both fully completed data cubes and the corresponding raw data are required. For this purpose, a number of different databases containing  are available. One major challenge for the task of snapshot mosaic HSI demosaicing using neural networks is the lack of real world ground truth data. 

The only databases available are those that use either the pushbroom method or spatio-spectral-line scans, usually realized by a liquid crystal. Hence, the data has different characteristics than snapshot mosaic data (e.g., missing cross-talk) and therefore a trained network cannot adequately represent reality. However, we are not aware of any data sets that include MSFAs. All HSI databases used in this work are presented in Tab.~\ref{tab:Databases} and include a large number of different colored objects.

One alternative to include real MSFA data into training is to downsample captured snapshot mosaic data as presented in \citet{dijkstra2019hyperspectral}. However, simple downsampling leads to differences in distances of adjacent pixels, which affects the network results in an unknown manner. Therefore, this approach is not been followed in this work.

\section{\uppercase{Network Architecture}} \label{sec:network}
In the following, three different architecture types are presented. All are based on different works recently published and have been modified in order to achieve improved demosaicing results.

\begin{table}[ht]
\caption{The number of parameters and sizes of the input images of the different networks. If two input sizes are stated, the network structure needs different input.}
\small
\centering
    {%\rowcolors{2}{black!80!white!20}{black!40!white!60}
    \begin{tabular}{r|ll}
    Network     & Parameter & Input Size\\ \hline \hline
    ID-ResNet-L & 697k & $[16\times 100 \times 100]$ \\ \hline
    ID-ResNet-S & 118k & $[16\times 100 \times 100]$ \\ \hline
    \multirow{2}*{ID-UNet} & \multirow{2}*{128k} & $[16\times 100 \times 100]$ \\
                               &                         & $[1\times 100 \times 100]$ \\ \hline
    \multirow{2}*{Parallel-S} & \multirow{2}*{331k} & $[16\times 100 \times 100]$ \\ %ParallelNet-Mod-3D
                               &                         & $[1\times 100 \times 100]$ \\ \hline
    \multirow{2}*{Parallel-L} & \multirow{2}*{382k} & $[16\times 25 \times 25]$ \\ %Parallel-ResNet
                               &                         & $[1\times 100 \times 100]$ \\ \hline
    \hline
    UNet        & 227k & $[1\times 100 \times 100]$ \\ \hline
    ResNet & \multirow{2}*{697k} & \multirow{2}*{$[16\times 100 \times 100]$} \\
    \citep{shinoda} &  &  \\ \hline
    Parallel            & \multirow{2}*{281k} & $[16\times 25 \times 25]$ \\ 
    \citep{wisotzky2022hyperspectral} &                         & $[1\times 100 \times 100]$ \\
    \end{tabular}}
    \label{tab:Networks}
\end{table}

\subsection{ResNet-based architecture}
In the initial phase, we aimed to enhance the network proposed by \citet{shinoda}. The modifications can be classified into two components: the preprocessing of input data and the CNN itself. Due to its superior effectiveness, we use ID interpolation as input of the network.

\begin{figure}[htp]
  \centering
  \includegraphics[width=\columnwidth]{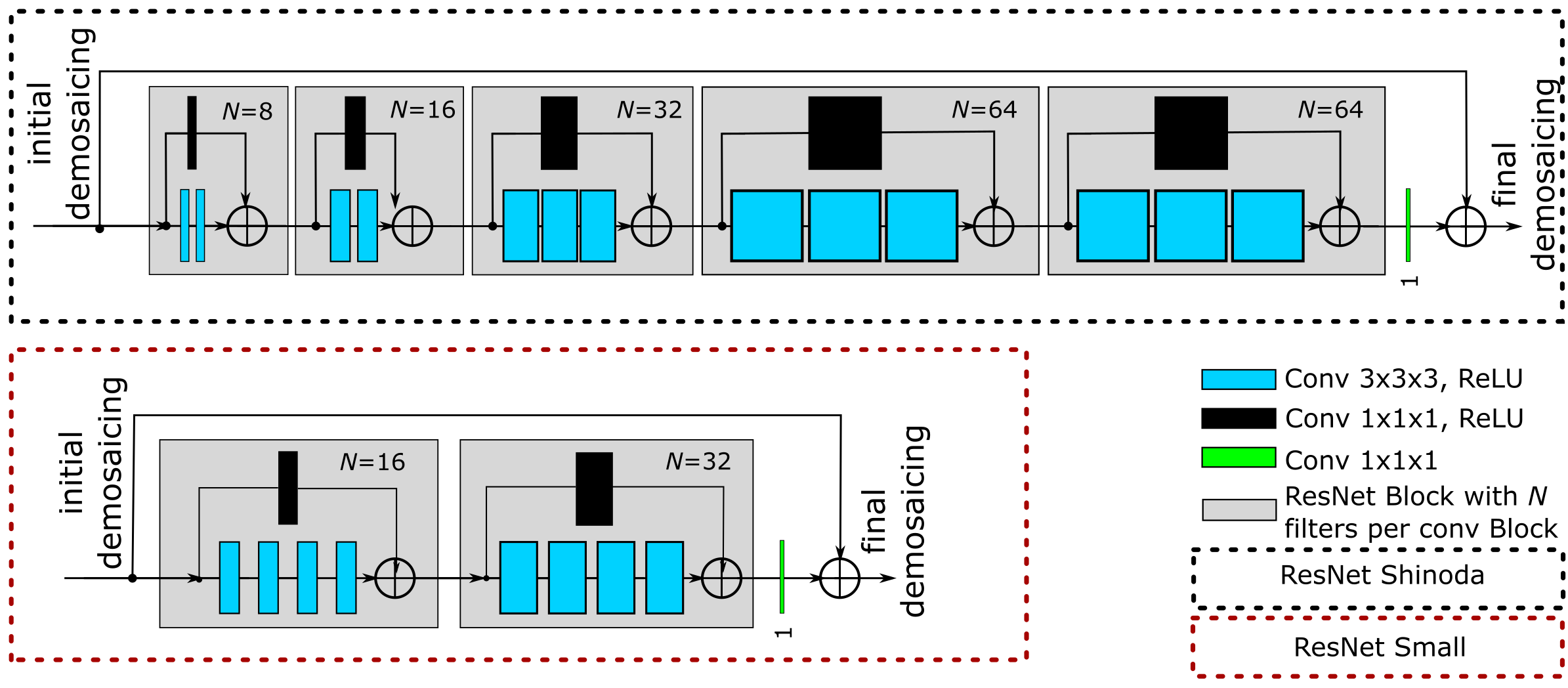}
  \caption{The ResNet-based architectures. Top: ResNet according to \citet{shinoda}. Bottom: our modified version.}
	\label{fig:ResNet}
\end{figure}

The alterations made to the CNN were intended to reduce the number of parameters without compromising the quality and are visualized in Fig.~\ref{fig:ResNet}. Initially, the network comprised of five residual blocks. The first two blocks contained two \textit{conv}-blocks in the main-path and one \textit{conv}-block in the skip-connection, while the next three residual blocks consisted of three \textit{conv}-blocks in the main-path and one \textit{conv}-block in the skip-connection. The number of filters increased from the first to the fifth residual block with 8/16/32/64/64, and were formed by a 3D kernel-size of $3\times3\times3$. Since our data input shows improved quality, we optimize the architecture by reducing the number of residual blocks to two. To achieve the desired quality of the CNN, we modified the number of \textit{conv}-blocks to four in the main-path and one in the skip-connection, with 16 filters in the first residual block and 32 in the second. 
These modifications reduced the overall number of parameters presented in Tab.~\ref{tab:Networks}.

\subsection{U-Net-based architecture}
\begin{figure}[htp]
  \centering
  \includegraphics[width=\columnwidth]{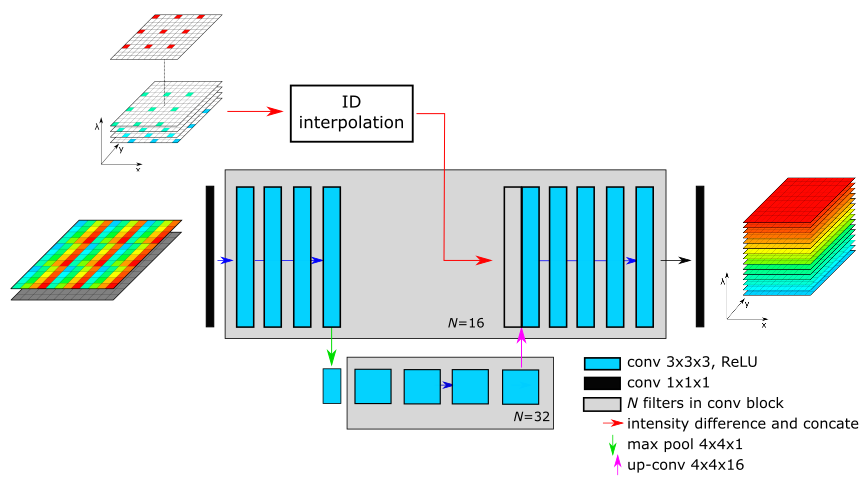}
  \caption{The U-Net-based architecture. A small U-Net structure is used, while instead of the skip connection, we insert the results from ID interpolation in the network.}
	\label{fig:UNet}
\end{figure}

Originally introduced for medical image segmentation, the U-Net is a classical approach to reconstruct images \citep{ronneberger2015u}. Specifically, the downsampling path of the U-Net captures the context of the image, while the upsampling path performs the complete reconstruction. To transfer information from the downsampling path to the upsampling path, skip connections are used. In this work, we modified the skip connections to insert external information for reconstruction improvements. We utilize a fully reconstructed image obtained by classical demosaicing, i.e., ID interpolation, which is then inserted into the upsampling path via the skip connection. A detailed schematic of this approach is provided in Fig.~\ref{fig:UNet}. 

The input to the U-Net is the 2D MSFA. To ensure a small, simple, and effective network architecture, we limit the downsampling layer to a max-pool $4\times4\times1$ operation and reconstruct the image in its 16 spectral regions using a $4\times4\times16$ kernel in the upsampling path. The \textit{conv}-blocks in the network consist of 16 filters in the upper layers and 32 filters in the lower layers and are provided with $3\times3\times3$ kernels.

\subsection{Parallel architecture}
\begin{figure}[htp]
  \centering
  \includegraphics[width=\columnwidth]{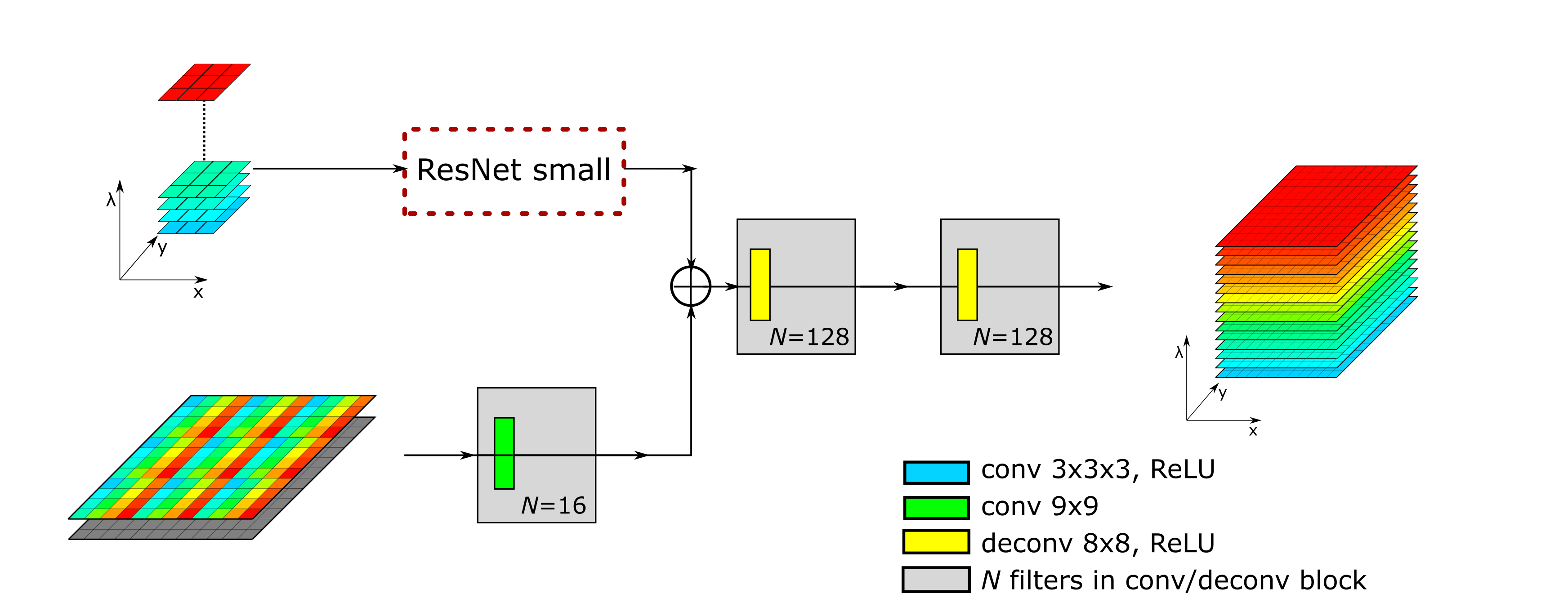}
  \includegraphics[width=\columnwidth]{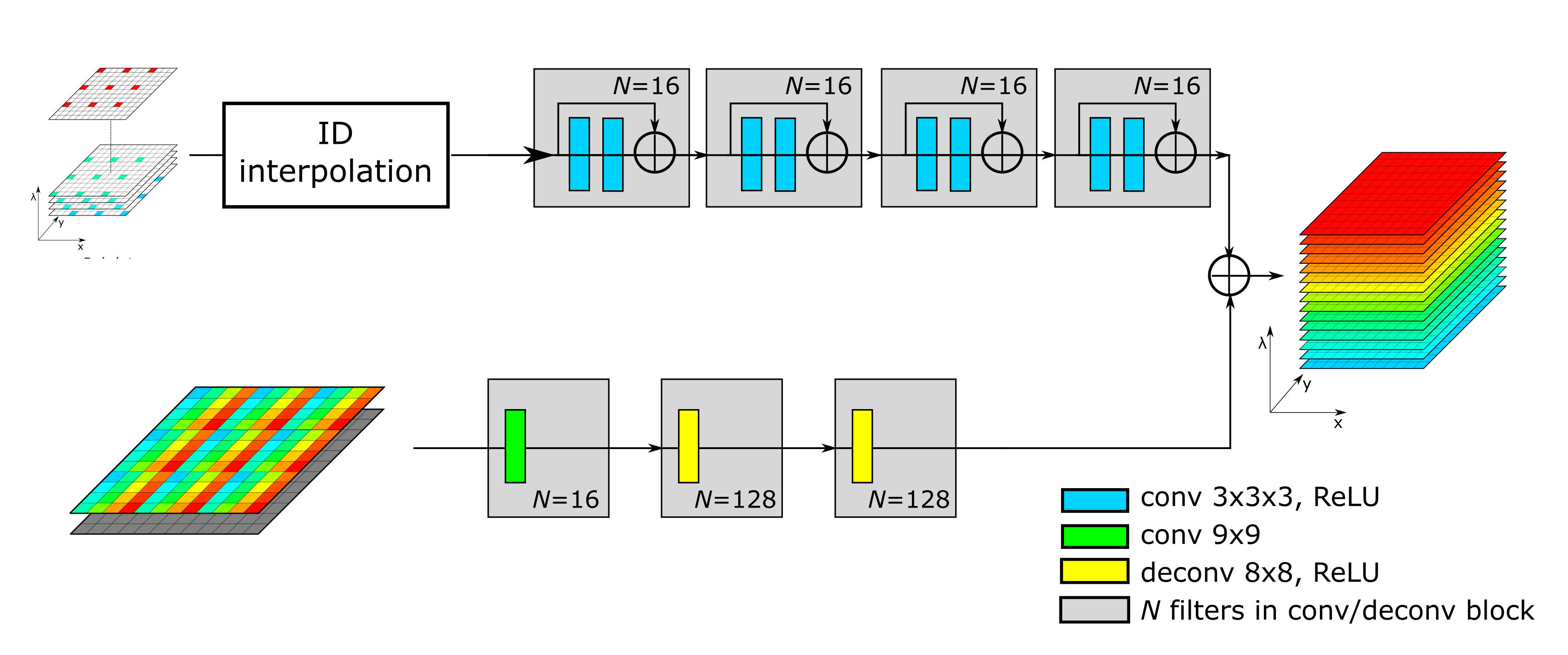}
  \caption{The parallel CNN architectures. Top: network Parallel-L is an extension of \citet{wisotzky2022hyperspectral} using the small ResNet-structure (cf.~Fig.~\ref{fig:ResNet}). Bottom: network Parallel-S is a combination of two effective approaches.}
	\label{fig:parallelMod}
\end{figure}

As third, we use a CNN architecture with parallel building blocks to reconstruct the correct spatial-spectral distribution in the image. We elaborated two general types based on literature \citep{dijkstra2019hyperspectral,shinoda,habtegebrial2019deep,wisotzky2022hyperspectral}.
First, two parallel feature extracting layers using a mosaic to cube converter (M2C) on one side and ResNet blocks on the other side are used followed by a feature adding and two deconvolution (\textit{deconv}) layers to upsample the spatial dimensions of the image. The second implementation combines two effective approaches introduced by \citet{dijkstra2019hyperspectral} and \citet{shinoda}/\citet{habtegebrial2019deep}, which are added and refined to form the demosaiced output.
Both implementations use 3D kernels and are presented in Fig.~\ref{fig:parallelMod}.

\section{\uppercase{Data and Training}} \label{sec:data_training}
During training, we used image patches of size $[1\times 100\times 100]$ (representing the mosaic sensor output) as input for the network. The output and ground truth data had a size of $[16\times 100 \times 100]$. Depending on the model, the input is either left unchanged or was transformed in the following way:
\begin{itemize}
    \item Transformation to a sparse 3D cube where empty values are filled with mean; size: $[16\times 100 \times 100]$
    \item Transformation to a 3D cube with low resolution and smaller spatial size; size: $[16\times 25 \times 25]$
    \item Transformation to a 3D cube by ID interpolation; size: $[16\times 100 \times 100]$.
\end{itemize}
On each side of the images, we removed four pixels resulting in a shape of $92\times 92$ px, because ID interpolation distorts those outermost pixels making it more difficult for the models to predict the correct interpolation.

In order to train and validate the networks proposed in this work, a large HSI data set was built from different data collections introduced earlier. To allow comparability between the individual data collections, the value range of all HSI data are normalized to $[0,1]$. %, including CAVE, TokyoTech [11], ICVL [12], HyTexiLa and google [14] data sets. 
The data set was split into $75\%$ training, $15\%$ validation and $10\%$ test data. 
As described, the data does not represent real snapshot mosaic behavior including cross-talk and characteristic filter responses for each mosaic pixel.
Therefore, an MSFA had to be simulated. First, we just selected 16 wavelengths in the range of $450-630$ nm, interpolated these bands from the HSI data and built an MSFA as widely been done in literature \citep{Arad_2020_CVPR_Workshops,Arad2022Challenge,gob2021multispectral,habtegebrial2019deep}. We refer to this data set as \textit{SimpleData}.

In addition, we built a data set better representing real captured MSFAs by transforming the stated reflectances of the data $r(\lambda)$ into real MSI snapshot camera measurements $r_b$ at band $b$ according to
\begin{equation}\label{eq:1}
    r_b = \frac{ \int^{\lambda_{\max}}_{\lambda_{\min}} T(\lambda) I(\lambda) f_b(\lambda) r(\lambda) d\lambda }{ \int_{\lambda_{\min}}^{\lambda_{\max}} T(\lambda) I(\lambda) f_b(\lambda) d\lambda},
\end{equation}
where $T$ is the optical transmission profile of the optical components of our used hardware setup, $f_b$ characterize the optical filter responses of each of the $b$ spectral bands of the used camera \citep{WisotzkyComparision2021,Wisotzky2020JMI}, and $I$ is the relative irradiance of the light source. All profiles are well known and allow to transform HSI data into real MSI snapshot camera outputs. We refer to this data set as \textit{SimRealData}.

For training, we used the ADAM optimizer with an adaptive learning rate strategy and an initial learning rate of $0.0002$. At every 10th epoch, the learning rate is reduced by the factor 0.9. We trained each model for 100 epochs. The batch size was $20$. The loss function for calculating the difference between the ground truth and the predicted full-spectrum hyperspectral cube is defined by the mean squared error (MSE)
\begin{equation}
MSE(o,p) = \frac{1}{N} \sum_{i=0}^{N}{|o_i-p_i|^2},
\end{equation}
where $o$ is the ground truth and $p$ is the predicted value.
To evaluate model performance, we calculated the structural similarity index (SSIM), peak signal-to-noise ratio (PSNR) and spectral angle mapper (SAM).

\section{\uppercase{Results and Discussion}} \label{sec:res}
In the following, we present the evaluation results of the predictions of our presented network architectures. For this purpose, they are examined and compared in a quantitative and qualitative manner.
We compare our proposed networks with classical interpolation approaches, bilinear and ID interpolation, as well as with different reference network approaches introduced in Sec.~\ref{sec:related_work}: ResNet \citep{shinoda}, HSUp \citep{dijkstra2019hyperspectral}, U-Net and Parallel \citep{wisotzky2022hyperspectral}.

Further, we analyze the resulting images visually and, to show the usability of our work, we visually analyze intraoperative snapshot images acquired during a surgery of parotidectomy. 

\subsection{Quantitative Results}
All networks, our proposed networks as well as the different references, were trained on the two created datasets \textit{SimpleData} and \textit{SimRealData}. For both dataset modifications, the networks learned to predict a full spectral cube of dimension $[16\times 100\times 100]$ from the given input images. Our networks outperformed or were in the same range as the reference networks for both datasets.

\subsubsection{\textit{SimpleData}}
\begin{table}[b]
\caption{Demosaicing results on \textit{SimpleData} patches. Best result is bold, second best is bold-italic and third rank is italic.}
\small
\centering
    \begin{tabular}{r|lll}
    Network     & SSIM            & PSNR [dB]        & SAM               \\ \hline \hline
    ID-ResNet-L & \textbf{0.9891} & \textbf{52.3846} & \textbf{3.48e-02} \\
    ID-ResNet-S &         0.9856  &         51.1377  &         3.90e-02  \\
    ID-UNet     & \textib{0.9868} & \textit{51.3331} & \textit{3.85e-02} \\
    Parallel-S  & \textit{0.9865} & \textib{51.4654} & \textib{3.78e-02} \\  %ParallelNet-Mod-3D
    Parallel-L  &         0.9857  &         51.1386  &         3.93e-02  \\ \hline %Parallel-ResNet
    
    UNet        &         0.9839  &         50.5408  &         4.25e-02  \\
    ResNet      &         0.9792  &         48.9982  &         4.07e-02  \\
    Parallel    &         0.9863  &         51.1316  &         3.96e-02  \\ %ParallelNet4
    HSUp        &         0.9846  &         50.8835  &         4.05e-02  \\ \hdashline
    Bilinear    &         0.9235  &         37.5282  &         5.90e-02  \\
    ID          &         0.9671  &         39.7182  &         5.27e-02  \\
    \end{tabular}
    \label{tab:ResultsSimple}
\end{table}
For the \textit{SimpleData} without cross-talk all modified and newly introduced networks perform better that the state-of-the-art methods, see Tab.~\ref{tab:ResultsSimple}.
The large modified ResNet-based model using ID interpolation as input performed significantly best. The larger the network in terms of trainable parameter the better the results for all three evaluation measures, but also the slower the network performance. In comparison with the reference ResNet model \citep{shinoda}, the quality of the input image, i.e., the quality of the initial demosaicing, is of high relevance.
Using better input quality (ID instead of bilinear interpolation), the ResNet is able to achieve an relatively higher increases in reconstruction accuracy.
It can be assumed that with even better initial demosaiced data by using network-based demosaicing methods, the performance is further improved. However, this would also increase the number of parameters and thus reduce performance.

The network Parallel-S follows the ID-ResNet-L model in reconstruction accuracy. It performed as second best in PSNR and SAM metrics and third in SSIM. The performance order is switched with the ID-UNet, which is third in PSNR and SAM, and second in SSIM.
The Parallel-L network, which includes the smaller ID-ResNet-S, shows an improvement in comparision to ID-ResNet-S. Both methods are fourth and fifth in terms of the analyzed metrics. Thus, all proposed methods and modifided networks performing better than the compared modalities from literature. This also shows that it is possible to greatly reduce the network complexity (by up to six times: 118k network parameters for ID-ResNet-S instead of 697k parameters for ID-ResNet-L) with only minor loss of quality compared to the best performing model.

Interestingly, also with respect to other recent work \citep{arad2022ntire}, it appears that the complexity of the models, i.e., the number of model paramter, has a great impact on the quality of the demosaicing results on simple, rather unrealistic, data sets. Because after all, the best results in this study are also achieved with the most complex models.

\subsubsection{\textit{SimRealData}}
\begin{table}[b]
\caption{Demosaicing results on \textit{SimRealData}. Best result is bold, second best is bold-italic and third rank is italic.}
\small
\centering
    \begin{tabular}{r|lll}
    Network     & SSIM            & PSNR             & SAM               \\ \hline \hline
    ID-ResNet-L & \textbf{0.9989} & \textbf{62.0690} & \textbf{1.29e-02} \\
    ID-ResNet-S &         0.9984  &         60.6000  &         1.55e-02  \\
    ID-UNet     & \textit{0.9985} &         60.7327  &         1.58e-02  \\
    Parallel-S  &         0.9983  & \textit{60.9021} & \textit{1.44e-02} \\ %ParallelNet-Mod-3D
    Parallel-L  &         0.9982  &         60.5968  &         1.52e-02  \\ \hline %Parallel-ResNet
    
    UNet        &         0.9973  &         58.3865  &         2.28e-02  \\
%    ResNet      & & & \\
    Parallel    & \textib{0.9986} & \textib{61.2197} & \textib{1.41e-02} \\ %ParallelNet4
    HSUp        &         0.9983  &         60.6398  &         1.50e-02  \\
%    Bilinear    &         0.9796  &         45.4610  &         3.16e-02  \\
%    ID          &         0.9890  &         48.1613  &         2.68e-02  \\
    \end{tabular}
    \label{tab:ResultsReal}
\end{table}
\begin{figure*}[!t]
  \centering
	\includegraphics[width=\textwidth]{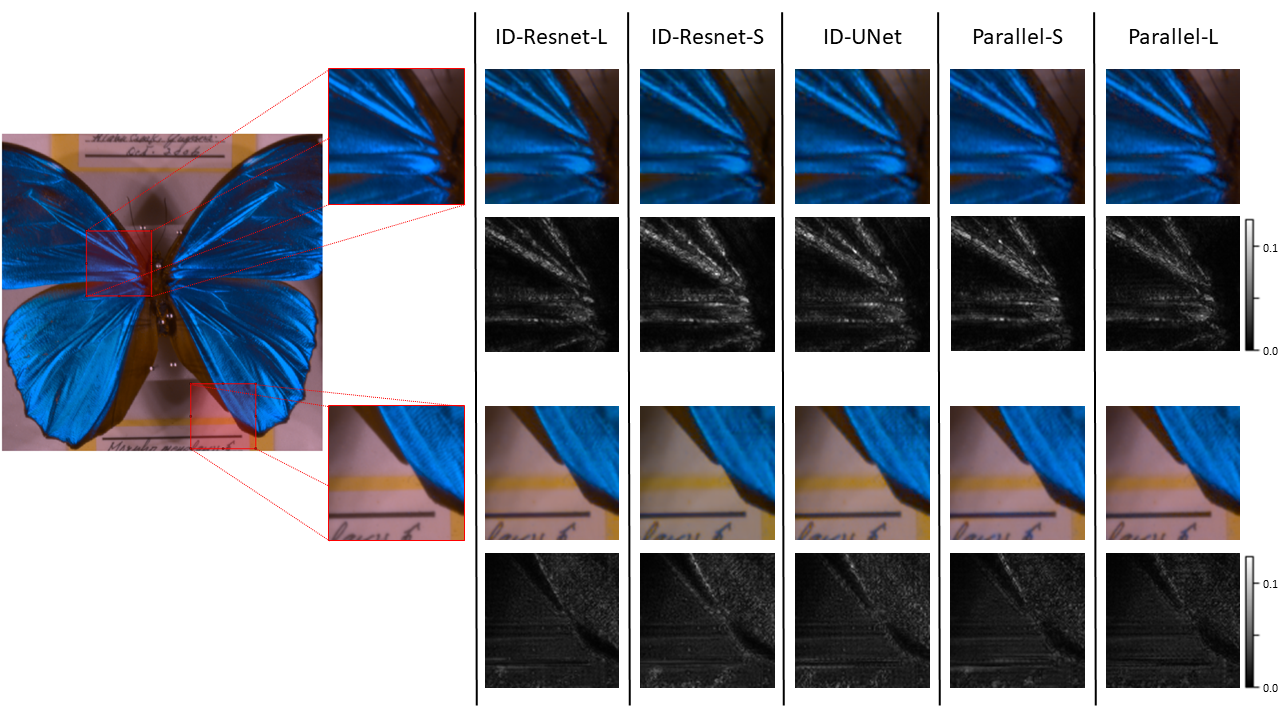} 
      \caption{Visual results and error images. Spectral data are represented in RGB and error images of two region of interest (ROI) are build using $l_1$-Norm. The maximum errors in the top ROI are 0.0598, 0.0994, 0.1001, 0.0805, 0.0667 and in the bottom ROI are 0.0401, 0.0597, 0.0614, 0.0520, 0.0366 in order of appearance of the models from left to right.}
	\label{fig:visres}
\end{figure*}
For the \textit{SimRealData}, which much better represent real captured data, the results of all networks, proposed as well as reference networks, are very much improved in comparison to \textit{SimpleData}, see Tab.~\ref{tab:ResultsReal}. Further, the difference between the reconstruction results of all networks is reduced. Again the largest network ID-ResNet-L performed best, but closely followed by the parallel networks. Interestingly, the state-of-the-art parallel model is performing second best, closely followed by Parallel-S. The other networks, expect for the standard U-Net, are following closely.
Thus, networks with only $40\%$ or $50\%$ of the parameters (Parallel or Parallel-S, respectively) achieve similarly good reconstruction results.

Obviously, the complexity of the networks is not decisive for the reconstruction results, but rather the quality of the input data and the general structure. The networks can draw information about all spectral bands from the complex spectral behavior of the individual pixels on an MSFA. The effect of cross-talk contains essential data, which are useful to all networks for a more precise spectral interpolation. This allows using simpler network structures while maintaining similar high reconstruction quality.

\subsection{Qualitative Results}
An analysis of the individual spectral channels did not reveal any discernible deviations in the quality characteristics between the spectral channels. Thus, no channel stands out as particularly defective during the demosaicing process. Therefore, the qualitative analysis of the results is made on RGB-calculated images.

\begin{figure}[t]
  \centering
	\includegraphics[width=\columnwidth]{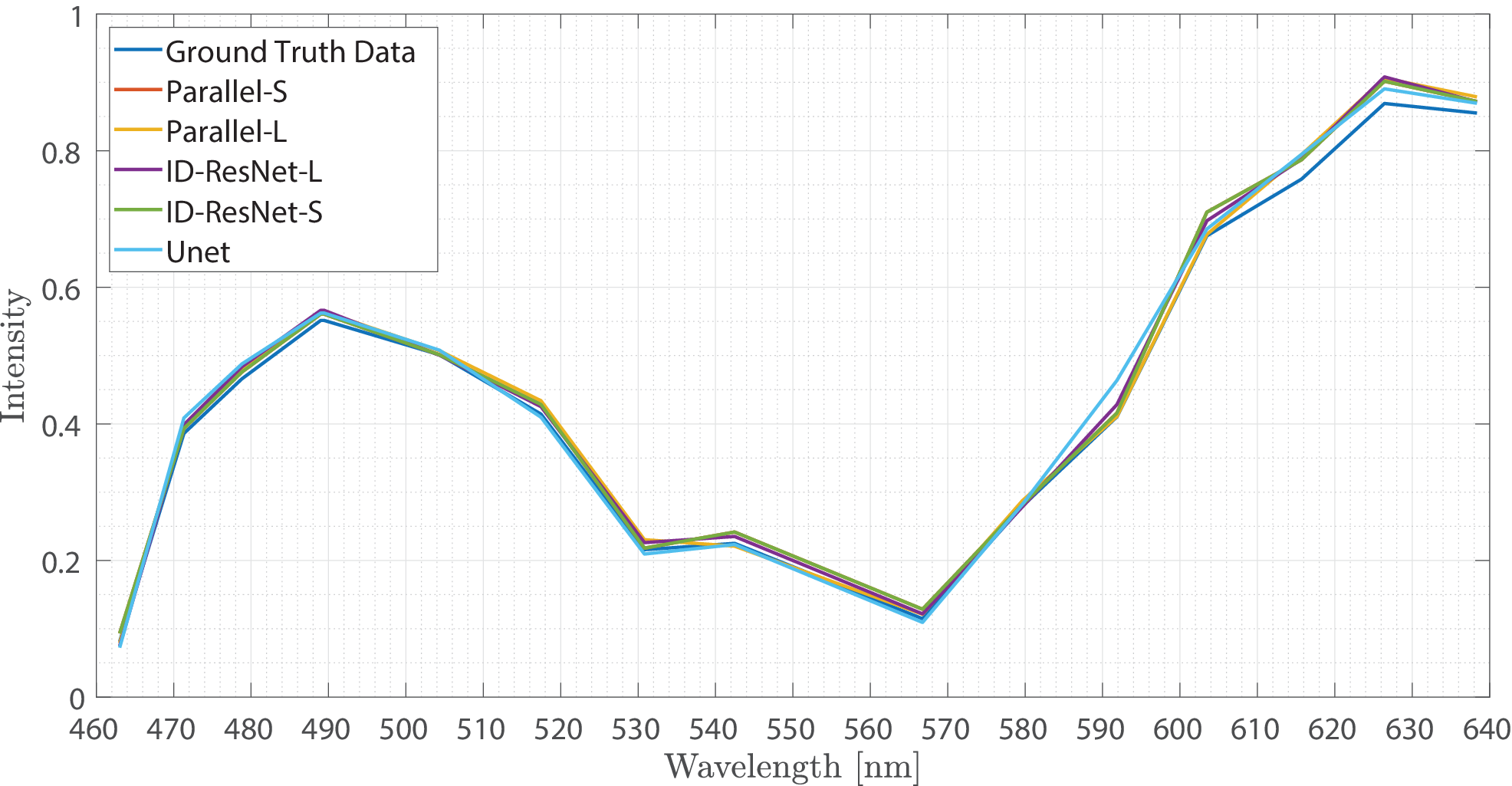}
  \caption{Spectral plot of a central left point in the left top wing of the butterfly in Fig.~\ref{fig:visres}. The SAM of the spectra in comparison to the ground truth are 0.0133, 0.0157, 0.0158, 0.0265, 0.0276 for the models ID-Resnet-L, Parallel-S, Parallel-L, ID-UNet and ID-Resnet-S, respectively.}
  \label{fig:plot}
\end{figure}

As can be seen in Fig.~\ref{fig:visres}, all reconstruction results appear in similar quality at first glance. On closer inspection, minor differences are evident at strong image edges, e.g., at the bottom right wing. These small differences can be quantitatively represented in a difference image (Fig.~\ref{fig:visres}) or in a spectral plot, Fig.~\ref{fig:plot}.

\begin{figure}[t]
  \centering
	\includegraphics[width=\columnwidth]{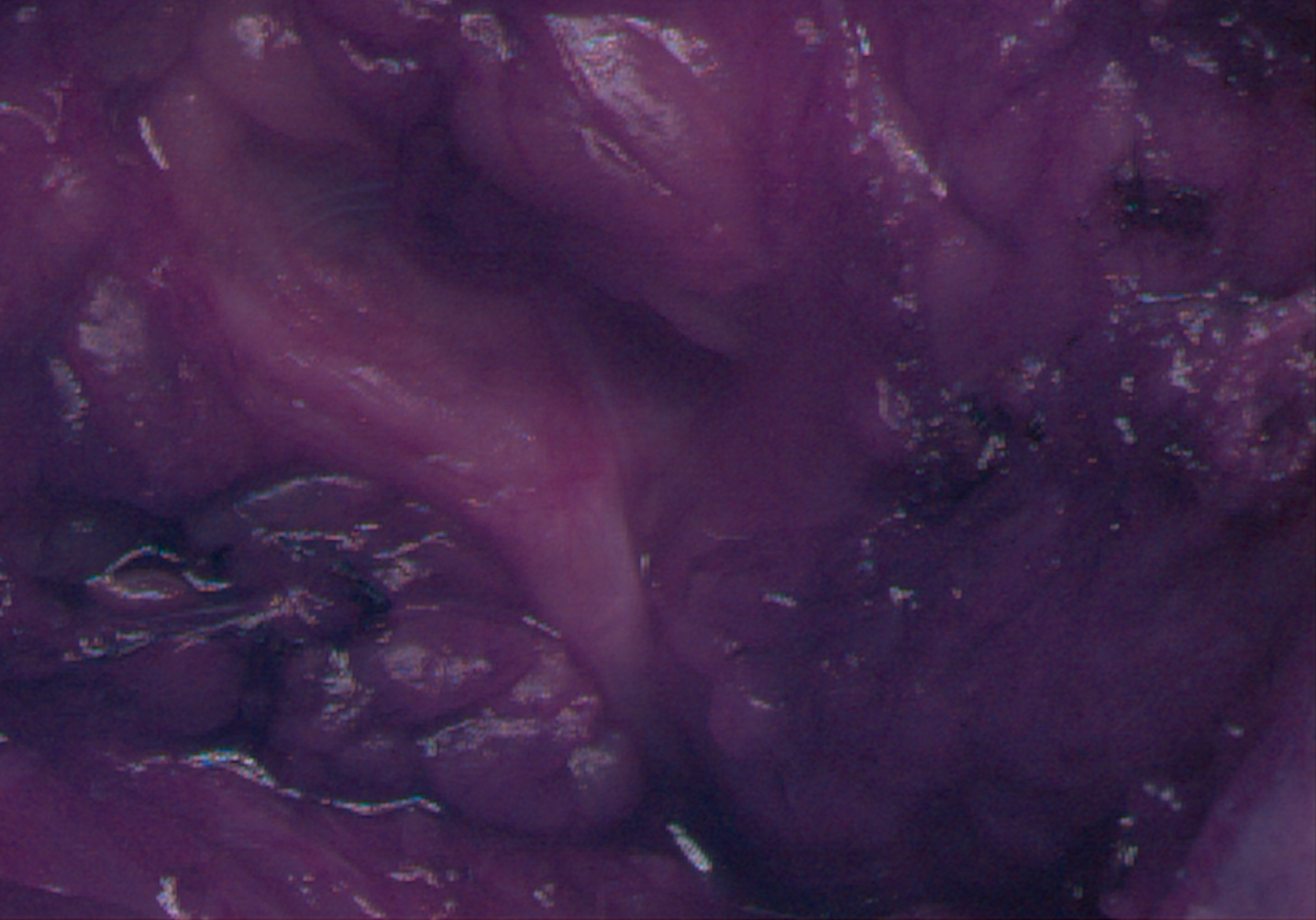}
  \caption{Demosaiced medical image showing qualitatively high image quality.}
  \label{fig:medima}
\end{figure}

In addition, we have demosaiced intraoperative images, see Fig.~\ref{fig:medima}. In terms of quality, these images show a high resolution as small details as well as edges are clearly visible. It is also noticeable that artifacts such as color fringes are reduced.

\section{\uppercase{Conclusion}}
\label{sec:discussion}
In conclusion, our results clearly show that significant improvements in demosaicing algorithms can be achieved by using well-designed neural network architectures. The networks and modifications we have introduced feature excellent reconstruction of the ground truth data while reducing or at least holding the model parameters constant. The results were tested both quantitatively and qualitatively, showing convincing results over traditional as well as CNN-based demosaicing methods.

Due to comparatively fewer network parameter, our networks result in more efficient computation proving the capability for real-time application, e.g., for intraoperative hyperspectral image application. Especially in comparison to current work \citep{arad2022ntire}, which uses very complex networks and achieves similar results, this is a gain. Importantly, our focus is on correct spectral reconstruction rather than visual attractiveness, which is of high importance for the mentioned applications and supported by the quantitative results confirmed by our qualitative evaluation.

Moreover, the differences between the results of different datasets, in agreement with the existing literature \cite{dijkstra2019hyperspectral}, provide valuable insights for the demosaicing of real camera data. This underscores the importance of developing demosaicing solutions that train on data being as close as possible to real MSFA data. Our results demonstrated that the use of synthetic representatives of real MSFA data are suitable for training and networks trained on these data perform well despite fewer training parameters, thereby enabling fast processing. This pursuit of efficient solutions is critical for practical applications in various fields, including medical imaging and remote sensing, allowing to integrate compact acquisition concepts like snapshot mosaic imaging into such processes.

\section*{\uppercase{Acknowledgment}}
This work was funded by the German Federal Ministry for Economic Affairs and Climate Action (BMWK) under Grant No.~01MK21003 (NaLamKI). Only tissue that has been exposed during normal surgical treatment has been scanned additionally with our described camera. This procedure has been approved by Charit\'e--Universit\"atsmedizin Berlin, Germany.

\bibliographystyle{apalike}
{\small
\bibliography{literature}}

\end{document}